\documentclass[a4paper,12pt]{article}
\usepackage[margin=1.3cm]{geometry}
\usepackage{comment}
\usepackage{amssymb,extarrows,graphicx,subfigure,setspace}
\usepackage{cite}
\usepackage{slashed}
\usepackage{tensor}
\usepackage[toc,page]{appendix}
\usepackage{color}
\usepackage{physics}
\usepackage{hyperref}
\usepackage{dirtytalk}
\hypersetup{colorlinks=true, linkcolor=blue, citecolor=red, linktoc=page}
\makeatother

\usepackage{float}
\usepackage{textcomp}
\usepackage{amsmath}
\newmuskip\pFqmuskip

\newcommand*\pFq[6][8]{%
  \begingroup 
  \pFqmuskip=#1mu\relax
  \mathcode`\,=\string"8000
  \begingroup\lccode`\~=`\,
  \lowercase{\endgroup\let~}\pFqcomma
  {}_{#2}F_{#3}{\left[\genfrac..{0pt}{}{#4}{#5};#6\right]}%
  \endgroup
}
\newcommand{\pFqcomma}{\mskip\pFqmuskip}

\usepackage{mathrsfs}
\usepackage{hyperref}

\newcommand{\be}{\begin{equation}}
\newcommand{\bea}{\begin{eqnarray}}
\newcommand{\eea}{\end{eqnarray}}
\newcommand{\ba}{\begin{array}}
\newcommand{\ea}{\end{array}}
\newcommand{\ee}{\end{equation}}
\newcommand{\bes}{\begin{equation*}}
\newcommand{\beas}{\begin{eqnarray*}}
\newcommand{\eeas}{\end{eqnarray*}}
\newcommand{\bas}{\begin{array*}}
\newcommand{\eas}{\end{array*}}
\newcommand{\ees}{\end{equation*}}

\numberwithin{equation}{section}

\begin{document}
\color{black}
\begin{center}
\Large{\bf The interplay of WGC and WCCC via charged scalar field fluxes in the RPST framework}\\
\small \vspace{0.7cm}{\bf Mohammad Reza Alipour$^{\star}$\footnote {Email:~~~mr.alipour@stu.umz.ac.ir}}, \quad
 \small \vspace{0.2cm} {\bf Jafar Sadeghi $^{\star}$\footnote {Email:~~~pouriya@ipm.ir}}, \quad\\\vspace{0.2cm}
{\bf Saeed Noori Gashti $^{\dag}$\footnote {Email:~~~saeed.noorigashti@stu.umz.ac.ir}},\quad
{\bf Mohammad Ali S. Afshar $^{\star}$\footnote {Email:~~~m.a.s.afshar@gmail.com}},\\
\vspace{0.5cm}$^{\star}${Department of Physics, Faculty of Basic
Sciences, University of Mazandaran\\ P. O. Box 47416-95447, Babolsar, Iran}\\
\vspace{0.3cm}$^{\dag}${School of Physics, Damghan University, P. O. Box 3671641167, Damghan, Iran}
\small \vspace{1cm}
\end{center}
\begin{abstract}
In this paper, we investigate the weak cosmic censorship conjecture (WCCC) for the Reissner-Nordström (R-N) AdS black hole within the framework of restricted phase space thermodynamics (RPST). Additionally, we consider energy flux and the equivalence mass-energy principle to examine the weak gravity conjecture (WGC) alongside the weak cosmic censorship conjecture. The interaction of incoming and outgoing energy flux leads to changes in the black hole. We examine whether the second law of thermodynamics holds in this scenario. It is noted that, when absorption and superradiance reach equilibrium, and by using the thermodynamics of black holes in restricted phase space, if the black hole is in or near an extreme state with radiation and particle absorption, the weak cosmic censorship conjecture is upheld. Furthermore, by employing the equivalence mass-energy principle and second-order approximation near extremality, we find that when the black hole radiates and its central charge exceeds the scaled electric charge, the superradiance particles adhere to the weak gravity conjecture. This causes the black hole to move away from its extreme state. However, when particles that obey the weak gravity conjecture are attracted to a very small black hole, the black hole approaches its extremality state.
\\

Keywords: Reissner-Nordström AdS Black Hole, Weak Gravity Conjecture, Weak Cosmic Censorship Conjecture, Restricted Phase Space
\end{abstract}
\tableofcontents
\section{Introduction}
Many attempts are made to establish a connection between quantum theory and gravity. Among these efforts is the swampland program, used for the phenomenology of quantum gravity (string theory).
In the swampland program, effective field theories compatible with quantum gravity are considered landscape and those not compatible are considered swampland \cite{19,20,21}. Also, one of the important conjectures of this swampland program, which defines the border between the landscape and the swampland, is the weak gravity conjecture, which is considered the upper limit of mass, hence gravity is the weakest force \cite{18,17}. Also,  different cosmological concepts such as inflation, dark energy, black holes physics, thermodynamics, etc investigated with the various conjectures of the swampland program. For further study, you can see in \cite{c,d,e,f,g,h,i,j,k,l,m,n,n',n'',p,p',q,r,s,t,u,v,w,x,y,z,aa,bb,cc,dd,ff,gg,ii,jj,kk,ll,mm,nn,oo,pp,qq,rr,ss,tt,uu,vv,ww,xx,yy,zz,aaa,bbb,ccc,ddd,eee,fff,ggg,hhh,iii}. One of the most important phenomena in cosmology is the black hole. Since the black hole is the solution to Einstein's equation, we can use the weak gravity conjecture derived from the string theory for the black hole to find a connection between these two theories.\\

Since general relativity breaks down near the fundamental singularity, Penrose proposed the weak cosmic censoring conjecture to solve this problem, where the singularity is hidden from distant observers by the black hole's event horizon \cite{17a,17b,8}. Therefore, people proposed experiments to validate the conjecture of weak cosmic censorship by absorbing or throwing up particles (the field). By doing this, they find out whether this event horizon remains a black hole or whether it is destroyed, in this case, the validity of the weak cosmic censorship conjecture is determined \cite{17c,17d,17e,17f,17g,17h,17i,17j,40,9,10,11}. Among these works, it can be mentioned \cite{40} that by using the thermodynamics of black holes in the extended phase space, they showed that for an extremal and near-extremal black hole despite the dispersion of charged fields in many times small,
stable black hole event horizon and weak cosmic censorship conjecture will hold. Note that they did the calculations by considering an isobaric process, i.e. $d\ell=0$. In this article, we examine the weak cosmic censorship conjecture by altering the phase space through restricted phase space thermodynamics, to determine if such changes impact the conjecture's validity.\\

The study of black hole thermodynamics in different phase spaces offers a rich landscape for exploring the fundamental aspects of gravitational physics. In extended phase space, thermodynamic variables are expanded to include the cosmological constant, treated as a dynamical pressure, which allows for a deeper understanding of the thermodynamic properties of black holes, especially in Anti-de Sitter (AdS) spacetimes. This framework has led to insights into the microstructure of black holes and the connection between classical thermodynamics and quantum physics \cite{40a,40b,40c}. Conversely, the restricted phase space approach focuses on a more constrained set of variables, often excluding the cosmological constant as a variable, thereby simplifying the thermodynamic study of black holes. This formalism is particularly useful in the context of holography, where it can reveal interesting behaviors of AdS black holes without the complexity introduced by the extended phase space variables \cite{1,2}.
The concept of restricted phase space thermodynamics (RPST) for AdS black holes via holography is a relatively new formalism proposed by \cite{1,2}. It's based on Visser's holographic thermodynamics but significantly modified \cite{41}: The AdS radius $(\ell)$ is held fixed as a constant. This means that the RPST does not involve the pressure \( P \) and volume \( V \) variables typically found in black hole thermodynamics. Instead, RPST introduces the central charge $(C)$ and chemical potential $\mu$ as a new pair of conjugate thermodynamic variables \cite{1,2,3,4,5,6,7,7a}. Recently, efforts have been made to establish the weak gravity conjecture and the weak cosmic censoring conjecture in various black holes. These findings demonstrate that the weak cosmic censoring conjecture remains valid for black holes even when $Q>M$ \cite{13,14,15,16,16a}. Now, in this article, we investigate the weak gravity conjecture for particles entering and leaving the black hole using the principle of equivalence of mass and energy to find their effect on the weak cosmic censorship conjecture.\\

So, the above information gives us motivation to arrange the paper as: In section 2 we show in extended phase space, the cosmological constant is equivalence to thermodynamic pressure and black hole's volume as its conjugate thermodynamic variable. Also, in this space, the mass of the black hole is regarded as enthalpy. Also in this section, we study the first law of thermodynamics and Smarr for the charged AdS black hole in the extended phase space.  By using the dual theory of holography, the central charge, also has a conjugate thermodynamic quantity as a chemical potential. We study the first law of thermodynamics and Smarr's relation for a charged AdS black hole in restricted phase space. In section 3, we generally examine the charged scalar field near the event horizon of the corresponding black hole. In this study, we examine the scattering behavior of the coupled massive scalar field possessing an electric charge interacting with the charged AdS black hole within a 4-dimensional spacetime framework.  The charged massive scalar field in charged Ads black hole is discussed in section 4.  In section 5,  we investigate  WCCC in near-extremal and extremal-charged AdS black holes with the above-mentioned information. In section 6 we study the WGC for the corresponding black hole with the help of the equivalence of mass and energy principle. Finally, we explain the conclusions in section 7.
\section{RPS thermodynamics and charged AdS black hole }
In the extended phase space, the cosmological constant $(\Lambda)$ is equivalence to the thermodynamic pressure of a black hole. As we know, the change of AdS radius $(\ell)$  will change the black hole's thermodynamic pressure, with the black hole's volume as its conjugate thermodynamic variable \cite{45},
\begin{equation}\label{eq1}
\begin{split}
P=-\frac{\Lambda}{8\pi}=\frac{(d-1)(d-2)}{16\pi \ell^2}, \qquad  V=\bigg(  \frac{\partial M}{\partial P} \bigg)_{S,Q},
\end{split}
\end{equation}
where $d$ and $r_h$ are the dimension and radius of the event horizon of the black hole, respectively.
Also, in this space, the mass of the black hole is regarded as enthalpy \cite{40,43}, and we have,
\begin{equation}\label{eq2}
\begin{split}
M=E+PV,
\end{split}
\end{equation}
where  $(E)$ is the internal energy of the black hole. In addition, the important relations of the first law of thermodynamics and Smarr for the charged AdS black hole  in the extended phase space can be expressed by the following expressions,
\begin{equation}\label{eq3}
\begin{split}
dM=TdS+\Phi dQ+VdP
\end{split}
\end{equation}
and
\begin{equation}\label{eq4}
\begin{split}
M=\frac{d-2}{d-3}TS+\Phi Q-\frac{2}{d-3}PV,
\end{split}
\end{equation}
here, $S$, $\Phi$, and $T$ are the entropy, electric potential, and Hawking temperature of the black hole respectively.
In restricted phase space, Newton's constant $(G)$ varies and the AdS radius $(\ell)$ is considered fixed, thus the black hole's mass is seen as its internal energy $(E)$ \cite{1,2,6,15},
\begin{equation}\label{eq5}
\begin{split}
E=M.
\end{split}
\end{equation}
By using the dual theory of holography, the central charge $(C)$ as an extensive variable has conjugate thermodynamic quantity as a chemical potential $(\mu)$ \cite{1,2,41}, 
\begin{equation}\label{eq6}
\begin{split}
C=\frac{\ell^{d-2}}{G},    \qquad   \mu=\bigg(  \frac{\partial E}{\partial C} \bigg)_{S,\hat{Q}},
\end{split}
\end{equation}
Also here, the  electric charge and corresponding potential are rewritten  by following equation,
\begin{equation}\label{eq7}
\begin{split}
\hat{Q}=Q\sqrt{C}, \qquad   \hat{\Phi}=\frac{\Phi(r_h)}{\sqrt{C}}=\frac{\hat{Q}}{r_h C}
\end{split}
\end{equation}
The first law of thermodynamics and Smarr's relation for a charged AdS black hole in restricted phase space are obtained by,
\begin{equation}\label{eq8}
\begin{split}
dE=TdS+\hat{\Phi} d\hat{Q}+\mu dC
\end{split}
\end{equation}
and
\begin{equation}\label{eq9}
\begin{split}
E=TS+\hat{\Phi} \hat{Q}+\mu C
\end{split}
\end{equation}
Next, we obtain the metric, entropy, mass, and temperature of the charged black hole in the restricted phase space. So, in that case, the metric background of the charged AdS black hole in 4-dimension will be the following form,
\begin{equation}\label{eq10}
\begin{split}
ds^2=-h(r)dt^2+\frac{1}{h(r)}dr^2+r^2d\theta^2+r^2 \sin^2\theta d\phi^2
\end{split}
\end{equation}
where \cite{1}
\begin{equation}\label{eq11}
\begin{split}
h(r)=1-\frac{2GM}{r}+\frac{GQ^2}{r^2}+\frac{r^2}{\ell^2} \qquad   or \qquad  h(r)=1-\frac{2\ell^2 M}{C r}+\frac{\ell^2 \hat{Q}^2}{C^2 r^2}+\frac{r^2}{\ell^2}.
\end{split}
\end{equation}
Here, we assume units as  $k_B=c=\hbar=1$. When $\frac{Q}{M}<\sqrt{G}$, the black hole has two internal and external event horizons, we denote the external event horizon by $r_h$. Also, in the case of $\frac{Q}{M}=\sqrt{G}$, the black hole is in an extreme state with only one event horizon.
Also, the entropy of the charged black hole in the corresponding  space is obtained by the following equation,
\begin{equation}\label{eq12}
\begin{split}
S=\frac{\pi r_h^2}{G}, \qquad  r_h=\ell\bigg( \frac{S}{\pi C} \bigg)^{\frac{1}{2}}
\end{split}
\end{equation}
By using equations \eqref{eq7},\eqref{eq11} and \eqref{eq12}, one can obtain the mass of  black hole as,
\begin{equation}\label{eq13}
M=\frac{\pi  C S+\pi ^2 \hat{Q}^2+S^2}{2 \pi ^{3/2} \ell \sqrt{C S}}
\end{equation}
Also, Hawking temperature and chemical potential are obtained by,
\begin{equation}\label{eq14}
T=\bigg(  \frac{\partial M}{\partial S} \bigg)_{\hat{Q},C}=\frac{\pi  C S-\pi ^2 \hat{Q}^2+3 S^2}{4 \pi ^{3/2} \ell S \sqrt{C S}}
\end{equation}
and
\begin{equation}\label{eq15}
\mu=\bigg(  \frac{\partial M}{\partial C} \bigg)_{S,\hat{Q}}=\frac{\pi C S-\pi ^2 \hat{Q}^2-S^2}{4 \pi ^{3/2} C \ell \sqrt{C S}}
\end{equation}
In the next section, we  are going to examine the charged scalar field near the event horizon of corresponding black hole.
\section{Charged massive scalar field in charged AdS black hole}
Black holes are capable of acquiring conserved quantities, including energy, momenta, and electric charge, through the interaction with an external field. The obtained conserved quantities from the black hole correspond to the fluxes of the external field. Through these fluxes, the black hole undergoes alterations in its state as it engages with the external field. By analyzing the fluxes, we can approximate the transformations occurring within the black hole over infinitesimal time intervals. In this study, we examine the scattering behavior of the coupled massive scalar field possessing an electric charge interacting with the charged AdS black hole within a 4-dimensional spacetime framework.
The equation of the massive charged scalar field near the charged black hole is given by \cite{40,43,44},
\begin{equation}\label{eq16}
\frac{1}{\sqrt{-g}}\partial_{\mu}(g^{\mu\nu}\sqrt{-g}\partial_{\nu}\Psi)-2iq g^{\mu\nu}A_{\mu}\partial_{\nu}\Psi-q^2 g^{\mu\nu}A_{\mu}A_{\nu}\Psi-\mu_s^2\Psi=0
\end{equation}
where $A^{\mu}=A^{\nu}=(\Phi(r),0,0,0)$, $g^{\mu\nu}$ , $\mu_s$,  $q$ and $\Psi$ are  the electric potential, black hole tensor metric, mass, electric charge and charged massive scalar field, respectively. To obtain the scalar field, we use separation of variables,
\begin{equation}\label{eq17}
\Psi(t,r,\theta,\phi)=e^{-i\omega t} R(r) Y_{l m}(\theta,\phi)
\end{equation}
where $R(r)$, $ Y_{l m}(\theta,\phi)$ and $\omega$  are radial functions, spherical harmonic functions, and energy of the particle, respectively.
Furthermore, the integer parameters $l$ and $m$ denote the spherical and azimuthal harmonic indices of the resonant eigenvalues that define the characteristics of the charged massive scalar fields within the charged black-hole spacetime. We get two radial and angular parts equations by putting relations \eqref{eq10} and \eqref{eq17} into equation \eqref{eq16},
\begin{equation}\label{eq18}
h(r)\frac{d}{dr}\bigg[r^2 h(r)\frac{dR(r)}{dr} \bigg]+\bigg[r^2\bigg(\omega-\frac{qQ}{r}\bigg) -h(r)\bigg(l(l+1)+\mu_s r^2\bigg) \bigg]R(r)=0
\end{equation}
\begin{equation}\label{eq19}
\bigg[\frac{1}{\sin^2\theta}\frac{\partial^2}{\partial\phi^2}+\frac{1}{\sin\theta}\frac{\partial}{\partial\theta}\bigg(\sin\theta\frac{\partial}{\partial\theta} \bigg) \bigg]Y_{lm}(\theta,\phi)=-l(l+1)Y_{lm}(\theta,\phi)
\end{equation}
To solve the radial equation, it is common practice to introduce the tortoise coordinate $\frac{dr}{dy}=h(r)$. In this case, the radial range of $r_h\leq r< +\infty$ becomes that of $-\infty<y\leq 0$ and the equation of the radial part becomes as,
\begin{equation}\label{eq20}
\frac{d^2R}{dy^2}+\frac{2h(r)}{r}\frac{dR}{dy}+\bigg[(\omega-\frac{qQ}{r})^2-\frac{h(r)}{r^2}\bigg( l(l+1)-\mu_s r^2\bigg) \bigg]R=0
\end{equation}
Since the flux of the scalar field into the black hole is important, we check equation \eqref{eq20} near the event horizon of the black hole $(r \rightarrow r_h)$, in which case we have a Schr\"{o}dinger-like equation,
\begin{equation}\label{eq21}
\frac{d^2R}{dy^2}+\bigg(\omega-\frac{qQ}{r_h}\bigg)^2R=0
\end{equation}
Therefore, the radial part of the scalar function is obtained,
\begin{equation}\label{eq22}
R(r)=e^{\pm i(\omega-\frac{qQ}{r_h})^2y}=e^{\pm i[\omega-q\Phi(r_h)]^2y}
\end{equation}
Here, the $+/-$ sign indicates the outgoing and ingoing of the fields.
Therefore, the wave function of the charged scalar field near the event horizon of the charged black hole takes the following form,
\begin{equation}\label{eq23}
\Psi(t,r,\theta,\phi)=e^{-i\omega t} e^{\pm i[\omega-q\Phi(r_h)]^2y} Y_{l m}(\theta,\phi)
\end{equation}
In this study, we examine alterations in a charged AdS black hole resulting from incoming/outgoing fluxes of the charged scalar field. As the scalar field enters/exits the black hole, changes in energy and electric charge will mirror those of the black hole itself. Consequently, variations in the properties of the black hole, contingent upon the energy and charge, are anticipated to adhere to a particular relationship. The energy and electric charge conveyed by the scalar field are represented by their fluxes at the outer horizon, which are derived from the energy-momentum tensor,
\begin{equation}\label{eq24}
T^{\mu}_{\nu}=\frac{1}{2}\mathcal{D}^{\mu}\Psi\partial_{\nu}\Psi^{*}+\frac{1}{2}\mathcal{D}^{*\mu}\Psi^{*}\partial_{\nu}\Psi-\delta^{\mu}_{\nu}\bigg( \frac{1}{2}\mathcal{D}_{\rho}\Psi \mathcal{D}^{*\rho}\Psi^{*}-\frac{\mu_s^2}{2}\Psi\Psi^{*}\bigg)
\end{equation}
with $D_{\mu}=\partial_{\mu}-iqA_{\mu}$. Also, the electric current of the charged scalar field is,
\begin{equation}\label{eq25}
j^{\alpha}=\frac{iq}{2}\Psi\mathcal{D}^{*\alpha}\Psi^{*}-\frac{iq}{2}\Psi^{*}\mathcal{D}^{\alpha}\Psi
\end{equation}
Therefore, from equations \eqref{eq24} and \eqref{eq25}, the energy flux and electric charge near the event horizon are obtained by \cite{40},
\begin{equation}\label{eq26}
\frac{dE}{dt}=\int \sqrt{-g} T^{r}_{t} d\phi d\theta=\omega(\omega-q\Phi_h)r_h^2
\end{equation}
\begin{equation}\label{eq27}
\frac{dQ}{dt}=-\int \sqrt{-g} j^{r} d\phi d\theta=\frac{q}{\omega}\frac{dE}{dt}=q(\omega-q\Phi_h)r_h^2.
\end{equation}
From the above equations, when $\omega > q\Phi_h$, the energy and charge flux are positive, indicating the particle's entry into the black hole. But for $\omega < q\Phi_h$, the energy and charge flux becomes negative, which means the exit of the particle from the black hole, and this phenomenon is called black hole superradiation. These fluxes can change the internal energy and electric charge of black holes in infinitesimal time interval $dt$.
Since we consider the thermodynamics of the black hole in the RPS, the changes in the internal energy of the black hole are equal to the changes in the mass of the black hole. Also, we consider the changes in electric charge according to equations \eqref{eq7} and \eqref{eq27}, which have been re-scaled in this space. So in that case, we have,
\begin{equation}\label{eq28}
dE=dM=\omega(\omega-\hat{q}\hat{\Phi}_h)r_h^2 dt
\end{equation}
\begin{equation}\label{eq29}
d\hat{Q}=\hat{q}(\omega-\hat{q}\hat{\Phi}_h)r_h^2dt+\frac{\hat{Q}}{2C}dC
\end{equation}
where
\begin{equation}\label{eq30}
\hat{q}=q\sqrt{C}, \qquad  \hat{\Phi}_h=\frac{\Phi_h}{\sqrt{C}}
\end{equation}
Since the charge in RPS is rescaled, the variations of charge changes in this space are different from the variations of charge in the normal and extended space. Because it depends on the central charge differential in addition to the time differential.
\section{Charged scalar field and RPS thermodynamics}
In RPS thermodynamics, the radius AdS is fixed, and Newton's constant is considered as a variable. It can also be found that by attracting a scalar field or a particle into the black hole, there are changes in the quantities $(M, \hat{Q}, C, r_h)$ of the black hole, which, according to relations \eqref{eq6} and \eqref{eq11}, also leads to changes in the metric of the black hole.
Also, WCCC, the event horizon of the black hole plays an important role and is obtained from the relation $h(M,\hat{Q},r_h,C)=0$.
Therefore, in the case where the black hole attracts the scalar field or particle, this event horizon must exist, in this case, the event horizon of the black hole changes and is obtained from the following relationship,
\begin{equation}\label{eq31}
h(M+dM,\hat{Q}+d\hat{Q},r_h+dr_h,C+dC)=0,
\end{equation}
and,
\begin{equation}\label{eq32}
\begin{split}
& h(M+dM,\hat{Q}+d\hat{Q},r_h+dr_h,C+dC)=h(M,\hat{Q},r_h,C)\\
& \qquad +\frac{\partial h}{\partial M}|_{r=r_h}dM+\frac{\partial h}{\partial \hat{Q}}|_{r=r_h}d\hat{Q}+\frac{\partial h}{\partial r}|_{r=r_h}dr+\frac{\partial h}{\partial C}|_{r=r_h}dC
\end{split}
\end{equation}
where
\begin{equation}\label{eq33}
\begin{split}
& h(M,\hat{Q},r_h,C)=0,  \qquad  \frac{\partial h}{\partial M}|_{r=r_h}=-\frac{2 \ell^2}{C r_h}, \qquad  \frac{\partial h}{\partial \hat{Q}}|_{r=r_h}=\frac{2 \ell^2 \hat{Q}}{C^2 r_h^2} ,\\
& \frac{\partial h}{\partial r}|_{r=r_h}=-\frac{2 \ell^2 \hat{Q}^2}{C^2 r_h^3}+\frac{2 \ell^2 M}{C r_h^2}+\frac{2 r_h}{\ell^2}=4\pi T,      \qquad \frac{\partial h}{\partial C}|_{r=r_h}=\frac{1}{C}\bigg(\frac{2 \ell^2 M}{C r_h}-\frac{2\ell^2 \hat{Q}^2}{C^2 r_h^2}\bigg)
\end{split}
\end{equation}
 By  using equations \eqref{eq28}, \eqref{eq29}, \eqref{eq30}, \eqref{eq31}, \eqref{eq31} and \eqref{eq33},  one can obtain the variations  in the event horizon,
\begin{equation}\label{eq34}
\begin{split}
dr_h=\frac{\ell^2 r_h (\omega-\hat{q} \hat{\Phi} )^2}{2 \pi  C T} dt-\frac{1}{4\pi C T}\bigg(1+\frac{r_h^2}{\ell^2} \bigg) dC
\end{split}
\end{equation}
Using equation \eqref{eq12}, we calculate the variations of  entropy,
\begin{equation}\label{eq35}
\begin{split}
dS=\frac{2 \pi  C r_h}{\ell^2} dr_h+\frac{\pi  r_h^2}{\ell^2} dC
\end{split}
\end{equation}
By combining equations \eqref{eq34} and \eqref{eq35}, we rewrite  the variations of entropy ,
\begin{equation}\label{eq36}
\begin{split}
dS=\frac{r_h^2 (\omega-\hat{q} \hat{\Phi} )^2}{T} dt-\frac{r_h }{2 \ell^2 T}\bigg( 1+\frac{r_h^2}{\ell^2}-2 \pi  r_h T\bigg) dC.
\end{split}
\end{equation}
When we set the central charge changes to zero, we observe entropy changes in normal space consistent with [32].
In addition, in the case that the temperature of the black hole will be as,
\begin{equation}\label{eq37}
\begin{split}
T=\frac{1}{2\pi r_h}\bigg(1+\frac{r_h^2}{\ell^2} \bigg).
\end{split}
\end{equation}
Despite maintaining the conditions of the RPS, the entropy changes will still reduce to the solution of the normal space.
According to equation \eqref{eq36}, we find that entropy changes can be negative; in this case, Newton's second law can be violated. If the Hawking temperature of the black hole has a lower bound, the entropy changes can always be positive,
\begin{equation}\label{eq38}
\begin{split}
T\geq\frac{1}{2\pi r_h}\bigg(1+\frac{r_h^2}{\ell^2} \bigg).
\end{split}
\end{equation}
Also, in the situation where $\omega=\hat{q} \hat{\Phi}$, and $\bigg(T<\frac{1}{2\pi r_h}\bigg[1+\frac{r_h^2}{\ell^2} \bigg]\bigg)$ the entropy changes are negative.

\section{WCCC in near-extremal and extremal cases}
The WCCC suggests that a black hole's singularity should be hidden by the horizon from an observer at infinity, preventing a naked singularity within the black hole's structure if it has a stable horizon. Our current study explores the stability of the outer horizon during the scattering of a charged scalar field, considering RPS thermodynamics. By absorbing the scalar field, the initial state $h(M, Q, r_h, C)$ transitions to the final state $h(M + dM, Q + dQ, r_h + dr_h, C+dC)$ over a small time interval $dt$. By setting the outer horizon as $h(M, Q, r, C) = 0$, we can determine the presence of the horizon by examining solutions in $h(M + dM, Q + dQ, r_h+dr_h, C+dC) = 0$. This analysis simplifies studying the change in the minimum value of $h$.

Initially, the minimum value of $h$ is negative or zero, indicating solutions corresponding to their respective horizons. If the scalar field fluxes enter the black hole, changes in the black hole's mass and charge occur due to the charged scalar field within $dt$, leading to variations in the minimum value based on these changes. If the minimum value becomes positive during these changes, no solution indicates a horizon in the final state, resulting in the black hole becoming a naked singularity, contradicting the WCCC. On the other hand, in other cases, the horizon consistently hides the interior of the black hole in the final state, supporting the validity of the WCCC. Thus, determining the sign of the minimum value in the final state is crucial for validating the WCCC in the context of the charged scalar field scattering. Our analysis focuses on determining this sign in the final state, which can be inferred from the initial state as the final state differs slightly due to conserved charges transferred through fluxes over $dt$.

This slight modification is more notable when the initial state pertains to a near-extremal or extremal black hole as opposed to a nonextremal one. The minimum value of a near-extremal black hole (including an extremal one) approaches zero. It may shift to a positive value due to minor alterations in the external scalar field. Therefore, our attention is directed towards a near-extremal black hole as the initial state, distinguished by the near-extremal circumstance at the minimum point $r_{min}$ with a negative constant $\delta \ll 1$ signifying the minimum value of $h$,

\begin{equation}\label{eq39}
\begin{split}
h_{min}\equiv h(r_{min})=1-\frac{2\ell^2 M}{C r_{min}}+\frac{\ell^2 \hat{Q}^2}{C^2 r_{min}^2}+\frac{r_{min}^2}{\ell^2},
\end{split}
\end{equation}

 \begin{equation}\label{eq40}
\begin{split}
\frac{\partial h_{min}}{\partial r_{min}}=-\frac{2 \ell^2 \hat{Q}^2}{C^2 r_{min}^3}+\frac{2 \ell^2 M}{C r_{min}^2}+\frac{2 r_{min}}{\ell^2}=0,
\end{split}
\end{equation}

 \begin{equation}\label{eq41}
\begin{split}
\frac{\partial^2 h_{min}}{\partial r_{min}^2}=\frac{6 \ell^2 \hat{Q}^2}{C^2 r_{min}^4}-\frac{4 \ell^2 M}{C r_{min}^3}+\frac{2}{\ell^2}>0.
\end{split}
\end{equation}
Therefore, by considering the fluxes that change the mass and electric charge of the black hole in infinitesimally small time changes $dt$, which lead to the change of position, the minimum location as $r_{min}+dr_{min}$ we have,
 \begin{equation}\label{eq42}
\begin{split}
h(M+dM, \hat{Q}+d\hat{Q}, C+dC , r_{min}+dr_{min})=h_{min}+\frac{\partial h_{min}}{\partial M} dM+\frac{\partial h_{min}}{\partial \hat{Q}} d\hat{Q}+\frac{\partial h_{min}}{\partial C} dC+\frac{\partial h_{min}}{\partial r_{min}} dr_{min},
\end{split}
\end{equation}
where
 \begin{equation}\label{eq43}
\begin{split}
& h_{min}=\delta \leq0, \qquad \frac{\partial h_{min}}{\partial M}=-\frac{2 \ell^2}{C r_{min}}, \qquad \frac{\partial h_{min}}{\partial \hat{Q}}=\frac{2 \ell^2 \hat{Q}}{C^2 r_{min}^2},  \\
&\frac{\partial h_{min}}{\partial C}=\frac{1}{C}\bigg(\frac{2 \ell^2 M}{C r_{min}}-\frac{2\ell^2 \hat{Q}^2}{C^2 r_{min}^2}\bigg),\qquad \frac{\partial h_{min}}{\partial r_{min}}=\frac{C^2 r_h r_{min} \left(\ell^2 r_h+r_h^3+2 r_{min}^3\right)-\ell^4 \hat{Q}^2 \left(2 r_h-r_{min}\right)}{C^2 \ell^2 r_h r_{min}^3}=0.
\end{split}
\end{equation}
By putting relations \eqref{eq28}, \eqref{eq29}, \eqref{eq30} and \eqref{eq34}  in relation  \eqref{eq42}, we get,
 \begin{equation}\label{eq44}
\begin{split}
&h(M+dM, \hat{Q}+d\hat{Q}, C+dC , r_{min}+dr_{min})=\\
&-\frac{2 \ell^2 r_h (\omega-\hat{q} \hat{\phi}_h )\big[r_h (\omega-\hat{q} \hat{\phi}_{min})-r_{min} (\omega-\hat{q} \hat{\phi}_h )\big]}{C r_{min}} dt+\frac{\left(r_h-r_{min}\right) \left[C^2 r_h r_{min} \left(\ell^2+r_h^2\right)-\ell^4 \hat{Q}^2\right]}{C^3 \ell^2 r_h r_{min}^2} dC.
\end{split}
\end{equation}

First, we examine the black hole in the  extremal state $r_{min}=r_ h$ and $h_{min}=\delta=0$, in this case, the above equation is rewritten as follows,
 \begin{equation}\label{eq45}
\begin{split}
h(M+dM, \hat{Q}+d\hat{Q}, C+dC , r_{min}+dr_{min})=0.
\end{split}
\end{equation}
Thus, despite the scattering, the black hole remains in an extremal state, and the WCCC is valid. Also, the result obtained in the RPS is the same as the EPS and the normal space.
To study the near-extremal black hole, we analyze the event horizon close to the minimum point of $f(r_{min})$, i.e. at $r_h = r_{min}+\epsilon$ and $0<\epsilon << 1$. In this scenario, equation \eqref{eq44} can be expressed as follows,
\begin{equation}\label{eq46}
\begin{split}
&h(M+dM, \hat{Q}+d\hat{Q}, C+dC , r_{min}+dr_{min})=\delta-\epsilon  \left[\frac{2 \ell^2  (\omega-\hat{q} \hat{\phi}_h ) (\omega-2 \hat{q} \hat{\phi}_h )}{C }+\mathcal{O}( \epsilon )\right] dt\\
&+ \epsilon  \left[\frac{C^2 r_h^2 \left(\ell^2+r_h^2\right)-\ell^4 \hat{Q}^2}{C^3 \ell^2 r_h^3}+\mathcal{O} (\epsilon) \right]dC.
\end{split}
\end{equation}
We also have,
\begin{equation}\label{eq47}
\begin{split}
h_{min}=\delta=-\epsilon\bigg[ \frac{C^2 \ell^2 r_h^2+3 C^2 r_h^4-\ell^4 \hat{Q}^2}{C^2 \ell^2 r_h^3}+\mathcal{O}( \epsilon )\bigg]
\end{split}
\end{equation}
Therefore, for small time changes that lead to small charge center changes, i.e. $dt\sim\epsilon$ and $dC\sim\epsilon$, equation \eqref{eq46} is obtained as follows,
\begin{equation}\label{eq48}
\begin{split}
h(M+dM, \hat{Q}+d\hat{Q}, C+dC , r_{min}+dr_{min})=\delta\pm\mathcal{O}( \epsilon ^2)<0.
\end{split}
\end{equation}
Hence, the adjustment to the minimum value can be considered negligible in the initial order of d. Consequently, it can be inferred that there is no alteration to the condition of the black hole: a nearly extremal black hole remains nearly extremal even with variations in mass and electric charge. Thus, the WCCC is valid for near-extremal black holes as well. However, according to relations \eqref{eq46} and \eqref{eq47}, we find that, in the condition that $dC>>1$, there is a possibility that the conjecture of weak cosmic censorship is violated. Also, when there are no changes in the central charge, the result obtained is in agreement with the result obtained in the extended phase space \cite{40}.
\section{WGC with equivalence of mass and energy}
Here we use the equivalence of mass and energy, and we have (with $\hbar=1$ and $c=1$),
\begin{equation}\label{eq49}
\begin{split}
E=\mu_s,  \qquad   E=\omega   \qquad   \Rightarrow   \qquad  \mu_s=\omega.
\end{split}
\end{equation}
Also, according to relation \eqref{eq7}, we can have,
\begin{equation*}\label{eq499}
q \Phi_h =\hat{q} \hat{\Phi}_h.
\end{equation*}
Now, considering the second order of $\epsilon$, we get some interesting results which is shown in Table 1.
\begin{center}
\begin{table}
  \centering
\begin{tabular}{|p{7cm}|p{3cm}|p{2cm}|p{4cm}|}
  \hline
  \hspace{2cm}Condition  & \hspace{0.7cm} fluxes  & \hspace{0.2cm} WGC & \hspace{1cm} State \\[3mm]
   \hline
 \hspace{1.1cm}$ \mu_s < q \Phi_h$  $\rightarrow$ $\frac{1}{\Phi_h}< \frac{q}{\mu_s}$ & Superradiance &   $C>>\hat{Q}$ & $ N-E \rightarrow N-E$\\[3mm]
   \hline
   $q\Phi_h< \mu_s <2 q \Phi_h$ $\rightarrow$ $\frac{1}{2\Phi_h}< \frac{q}{\mu_s}< \frac{1}{\Phi_h}$ & Absorption & $C>>\hat{Q}$ , $r_{min}=\epsilon$ & $ N-E \rightarrow S-E$  \\[3mm]
  \hline
  \hspace{2cm}$ \mu_s >2 q \Phi_h$   & Absorption &\hspace{0.75cm} $ \times$ &  $ N-E \rightarrow N-E$ \\[3mm]
  \hline

\end{tabular}
\caption{Summary of the results}\label{4}
\end{table}
 \end{center}
Using the equivalence of mass and energy and relation \eqref{eq46}, we find that the black hole emits particles in its radiation that obey the weak gravity conjecture, which nevertheless holds the weak cosmic censorship conjecture so that the black hole from The extremity becomes more distant. But in the case where the black hole attracts the particle, two situations occur, in the first case, the black hole attracts the particle that follows the WGC. In this case, the WCCC is valid, with the difference that the black hole is closer to the extremity state.
In the second case, the black hole attracts a particle that does not follow the WGC, the WCCC will be valid so that the black hole moves away from the extremity state.
Next, we obtain the electric potential of the black hole near the extreme state. Given $r_h=r_{min}+\epsilon$, the electric potential can be expressed as,
\begin{equation}\label{eq50}
\begin{split}
\Phi_h=\frac{Q}{r_h}=\frac{Q}{r_{min}+\epsilon}=\frac{\hat{Q}}{r_{min}\sqrt{C}}(1-\frac{\epsilon}{r_{min}}).
\end{split}
\end{equation}
Also, by using $h(r)=0, T=\frac{h^{\prime}(r)}{4\pi}$ and 11, we can obtain $r_{min}$ as follows,
\begin{equation}\label{eq51}
\begin{split}
r_{min}=\frac{\sqrt{G}}{\sqrt{6}}\bigg(\sqrt{C^2+12 \hat{Q}^2}-C\bigg)^{\frac{1}{2}}.
\end{split}
\end{equation}
By placing \eqref{eq50} and \eqref{eq51} into Superradiance  conditions ( $\frac{1}{\Phi_h}< \frac{q}{\mu_s}$), we have,
 \begin{equation}\label{eq52}
\begin{split}
\frac{q}{\mu_s}>\sqrt{G}\big(1+\frac{\epsilon}{r_{min}} \big)\frac{\sqrt{C}\big(\sqrt{C^2+12 \hat{Q}^2}-C\big)^{\frac{1}{2}}}{\sqrt{6}\hat{Q}}.
\end{split}
\end{equation}
According to relation \eqref{eq52}, in the condition that $C>>\hat{Q}$, we have $\frac{q}{\mu_s}>\sqrt{G}\big(1+\frac{\epsilon}{r_{min}}\big)$, in which case the particles emitted from the black hole follow the WGC. But in the case where the black hole absorbs the particle, even when $C>>\hat{Q}$, we have,
  \begin{equation}\label{eq53}
\begin{split}
\frac{\sqrt{G}}{2}\big(1+\frac{\epsilon}{r_{min}} \big)<\frac{q}{\mu_s}<\sqrt{G}\big(1+\frac{\epsilon}{r_{min}} \big).
\end{split}
\end{equation}
Therefore, the WGC does not hold in this case. The black hole attracts particles that do not obey the WGC and repels particles that obey the WGC. In both cases the WCCC holds.
Also, according to relation \eqref{eq53}, we find that only if the black hole has a very small event horizon and is in an extreme state $(r_{min}=\epsilon)$ can absorb particles that follow the WGC. Therefore, these small black holes get closer to their extreme limit, which increases the probability of the black hole collapsing.
\section{Conclusions}
We employed the RPS thermodynamics and investigated the  WCCC. Also in this paper, we used the equivalence of mass and energy principle and discussed the relationship between two important conjectures WCCC  which is derived from general relativity, and WGC is derived from string theory. Here, we assumed that the internal charge and energy of the black hole change with the superradiance and absorption of the particle (or scalar field) in an infinitesimal time interval. By using the reproduced first law of thermodynamics, we have shown that the second law holds in conditions where the temperature has a lower limit as $(T\geq\frac{1}{2\pi r_h}\bigg(1+\frac{r_h^2}{\ell^2} \bigg))$. But the second law of thermodynamics can be violated in conditions where the temperature has an upper limit and the black hole is in equilibrium $\omega=\hat{q} \hat{\Phi_h}$.
Also, here we point out that when the central charge does not change, the obtained results are the same as the results obtained in normal space. In addition, we showed using the RPS thermodynamics that the WCCC holds despite the input and output flux in the extremal state and close to the extremal state. Therefore, the result obtained in the RPS is consistent with the results obtained in the normal space and the extended phase space(EPS). By applying the equivalence mass and energy principle for the particle and second-order approximation in the state close to the extreme black hole, we find that when we have superradiance, particles are emitted that obey the WGC, of course, with the condition that the $C>>\hat{Q}$. In this case, the black hole will be further away from its extreme state. Also, a black hole will only attract a particle that obeys the WGC if the black hole is very small and the central charge is larger than the scaled electric charge $C>>\hat{Q}$. In this case, the black hole becomes closer to its extreme state.

\end{document}